\documentclass[12pt,preprint]{aastex}

\begin{document}
\title{Metal Enrichment of the Intergalactic Medium and
Production of Massive Black Holes}
\author{Y.-Z. Qian\altaffilmark{1} and G. J. Wasserburg\altaffilmark{2}}
\altaffiltext{1}{School of Physics and Astronomy, University of
Minnesota, Minneapolis, MN 55455; qian@physics.umn.edu.}
\altaffiltext{2}{The Lunatic Asylum, Division of Geological and
Planetary Sciences, California Institute of Technology, Pasadena, CA
91125.}

\begin{abstract}
A model for the chemical evolution of the intergalactic medium (IGM) 
is presented using theoretical yields of very massive 
($M_{\rm VMS}>100\,M_\odot$) stars (VMSs) and Type II supernovae 
(SNe II). It is shown that if [Si/C] is indeed as high as $\sim 0.7$
in the IGM, then VMSs 
($M_{\rm VMS}\approx 140$--$260\,M_\odot$)
associated with pair-instability supernovae (PI-SNe) in low-mass
($\sim 10^5\,M_\odot$) halos at high redshift must produce at least 
50\% of the Si. The 
remainder is from later galactic outflows of SN II debris, which 
also provide most of the C and O. Both sources are required to
account for the metal inventory in the IGM.
The early VMS production must 
continue until redshift $z\sim 15$ so that the efficiency of VMS 
formation per low-mass halo is significantly below unity. 
Contributions from the later galactic outflows mainly 
occur at $z\sim 4$--6. Using a Salpeter initial mass function,
we infer that the number of VMSs 
($M_{\rm VMS}\approx 260$--$2000\,M_\odot$) producing
massive black holes (MBHs) with an average mass 
$\langle M_{\rm MBH}\rangle\sim 270$--$550\,M_\odot$ is
$\approx 0.72$ times the the number of VMSs associated with PI-SNe.
The amount of metals (particularly Si) in the IGM that is 
attributable to PI-SNe is thus closely coupled with the total mass 
of MBHs produced in epochs prior to galaxy formation. Production
of $\sim 50$\% of the Si in the IGM by PI-SNe corresponds to an early
inventory of MBHs that constitutes a fraction 
$\sim (4$--$8)\times 10^{-5}$ of the total baryonic mass in the 
universe. This is comparable to the global mass budget of the 
central supermassive black holes (SMBHs) in present-day galaxies.
The corresponding occurrence rates in each halo of 
$\sim 10^5\,M_\odot$ during the epoch of VMS formation at $z\gtrsim 15$
are $\sim 0.9$~Gyr$^{-1}$ for VMSs associated with PI-SNe and 
$\sim 0.6$~Gyr$^{-1}$ for the concomitant more massive stars
producing MBHs. These rates may be 
of use to studies of H$_2$ dissociation and reionization and to
models of SMBH formation.
\end{abstract}


\section{Introduction}

In this paper we explore implications of data
on C, O, and Si in the intergalactic medium (IGM) for the 
``metal'' sources. In earlier works (Wasserburg \& Qian 2000; 
Qian \& Wasserburg 2002; see also Qian, Sargent, \& Wasserburg 2002),
we proposed that very massive ($M_{\rm VMS}>100\,M_\odot$) stars (VMSs) 
formed from big bang debris produced a prompt metal inventory in the 
IGM at the level of [Fe/H]$_{\rm IGM}\equiv\log{\rm (Fe/H)}_{\rm IGM}-
\log{\rm (Fe/H)}_\odot\sim -3$, where (Fe/H) is the number ratio of Fe 
to H. This was based on observations of metal-poor stars in the Galactic
halo that showed a sudden jump in the production of heavy
rapid-neutron-capture ($r$-process) elements such as Ba and Eu
relative to Fe at [Fe/H]~$\sim -3$ (e.g., McWilliam et al. 1995;
Burris et al. 2000). We inferred that the jump was due to the onset 
of major production of the heavy $r$-process elements by a class of 
supernovae associated with normal stars, which could predominantly
form only after VMSs had provided a metallicity at the level of
[Fe/H]$_{\rm IGM}\sim -3$ to the IGM. This is supported by theoretical 
models of star formation, which suggested that a critical
metallicity of $\sim 5\times 10^{-4}$ times the solar
value is required for sufficient cooling of gas clouds to form 
low-mass protostellar aggregates (Bromm et al. 2001; see also
Bromm \& Loeb 2003). Using VMS yields of Heger \& Woosley (2002,
hereafter HW02), Oh et al. (2001) showed that there would be 
sufficient hard UV photons emitted from VMSs to reionize the
universe if these stars were to provide [Si/H]$_{\rm IGM}\approx -2.3$
(e.g., Cowie \& Songaila 1998; Ellison et al. 2000; see also 
Aguirre et al. 2004). We here consider both early VMSs
and later galactic outflows for metal enrichment of the IGM. 
It is argued that VMSs ($M_{\rm VMS}\approx 140$--$260\,M_\odot$)
associated with pair-instability supernovae (PI-SNe; HW02) are 
the dominant source of Si and also made
substantial contributions to C and O. The dominant contributions 
to C and O are from later galactic outflows governed by Type II 
supernovae (SNe II). We then show that for a plausible initial
mass function (IMF), the metal (particularly Si) inventory in the IGM
due to PI-SNe is coupled with production of black holes by
VMSs with $M_{\rm VMS}>260\,M_\odot$.
Thus, there is concomitant production of both metals and massive black 
holes at early epochs.

The present study is an extension of our previous work (Qian \&
Wasserburg 2005, hereafter QW05), where we sought to explain the
data on C, O, and Si in the IGM using a number of VMS and SN II models
and considering a wide range of IMFs for VMSs. In that work, we showed 
from detailed arguments that VMSs associated with PI-SNe are required 
to account for the Si/C ratio in the IGM and that the preferred scenario 
for metal enrichment of the IGM is a combination of contributions from
early VMSs and later galactic outflows. In the present paper, we assume 
the basic conclusions of that work and adopt the same input and 
formalism to develop the general
self-consistent implications of the chemical evolution model used 
there. The new results to be presented include the constraint on the
redshift for termination of VMS contributions and the relationship 
between metal production by PI-SNe and the early inventory of massive 
black holes.

We first review the issues.
There is a substantial inventory of metals in the general IGM 
over a wide range of redshift, $z=1.5$--5.5 (e.g., Songaila 2001; 
Pettini et al. 2003). Detailed studies of quasar absorption spectra
have established abundances of the ionic species \ion{C}{3}, \ion{C}{4}, 
\ion{O}{6}, \ion{Si}{3}, and \ion{Si}{4} in the Ly$\alpha$ forest (see Schaye 
et al. 2003; Aguirre et al. 2004; and Simcoe et al. 2004 for recent
analyses). The corresponding elemental abundances have been determined 
using models of the UV background (UVB; Haardt \& Madau 1996, 2001) to 
account for the fraction of an element in a 
specific ionization stage. Two typical UVB models are model Q, which
has a harder UVB provided by quasars only, and model QG, which has a 
softer UVB provided by both quasars and galaxies. For both models,
Schaye et al. (2003) and Aguirre et al. (2004) found that 
C and Si abundances are quite variable over different regions. 
For a given density of the Ly$\alpha$ forest, the abundance of,
say C, appears to follow a lognormal distribution.  
The net inventory of C over the density range (0.32 to 63 times 
the cosmic mean) covered by the data is
[C/H]$_{\rm IGM}=-2.3$ for UVB model Q and $-2.8$ for model QG.
For either model, Schaye et al. (2003) and 
Aguirre et al. (2004) found that neither [C/H]$_{\rm IGM}$ nor
[Si/H]$_{\rm IGM}$ show evidence of evolution in the range
$z=1.8$--4.1. Using the data at $z\sim 2.5$ and similar UVB models,
Simcoe et al. (2004) also found lognormal distributions for
C and O abundances. The results of the above groups
for [C/H]$_{\rm IGM}$ are in good agreement for either UVB model.
The observed variability of abundances in different regions 
must be a result of varying degrees of mixing.
However, there is no apparent variation of the Si/C or C/O ratio
(Aguirre et al. 2004; Simcoe et al. 2004).
In the models presented below, we will focus on the net 
IGM inventory and the associated elemental ratios.
The distribution of metals is not treated. 

From the observations there is no apparent evolution of the IGM 
inventory over $z=1.8$--4.1. Thus, this inventory must have been 
produced prior to $z\sim 4$. The plausible sources are 
VMSs and SNe II. The Si/C and C/O ratios\footnote{For both the 
IGM data and the
results from stellar models, we use the same reference solar
abundances as in QW05: $\log\epsilon_\odot({\rm Si})\equiv
\log({\rm Si/H})_\odot+12=7.55$ (Anders \& Grevesse 1989) adopted 
by Aguirre et al. (2004), and $\log\epsilon_\odot({\rm C})=8.52$
and $\log\epsilon_\odot({\rm O})=8.83$ (Grevesse \& Sauval 1998)
adopted by Simcoe et al. (2004). As Schaye et al. (2003) and
Aguirre et al. (2004) adopted a slightly different solar C
abundance of $\log\epsilon_\odot({\rm C})=8.55$ 
(Anders \& Grevesse 1989), we have made appropriate small
adjustments to their [C/H]$_{\rm IGM}$ and [Si/C]$_{\rm IGM}$
values for consistency.}
are potential diagnostics of the sources. These ratios 
are sensitive to the UVB model. 
As both of these ratios are available for models Q and QG, we
discuss these two UVB models first. While model Q gives
[Si/C]$_{\rm IGM}=1.45$ and [C/O]$_{\rm IGM}=0$, model QG gives
much lower values of [Si/C]$_{\rm IGM}=0.74$ and 
[C/O]$_{\rm IGM}=-0.50$ (Aguirre et al. 2004; Simcoe et al. 2004).
For both models [Si/C]$_{\rm IGM}$ is quite high and points to VMS
sources (Schaerer 2002; Aguirre et al. 2004). 
For UVB model QG, [Si/C]$_{\rm IGM}$ is well above
the yield ratio of SN II sources but below that of pure VMS 
sources (see Table 2 of QW05). 
The value of [Si/C]$_{\rm IGM}$ for model Q is so high
that it can be barely matched by the yield ratio of a pure VMS 
source. We have shown earlier (QW05) that the values of 
[Si/C]$_{\rm IGM}$ and [C/O]$_{\rm IGM}$ 
for UVB model Q cannot be accounted for simultaneously by any 
existing stellar models. However, both elemental ratios for
model QG can be matched using a combination of VMS and SN II 
sources. Results on [Si/C]$_{\rm IGM}$ only were also given by
Aguirre et al. (2004) for models QGS and QGS3.2, both of which
have an even softer UVB than model QG. Model QGS gives
[Si/C]$_{\rm IGM}=0.23$ while model QGS3.2 gives 
[Si/C]$_{\rm IGM}=0.43$. These [Si/C]$_{\rm IGM}$ values are
consistent with the yield ratio of SN II sources but significantly below
that of pure VMS sources  (see Table 2 of QW05).
Consequently, were the composition of the IGM
characterized by these [Si/C]$_{\rm IGM}$ values, there would be
no requirement of VMS contributions. However,
Schaye et al. (2003) showed that for UVB model QGS, the C
abundance strongly decreases with time for all densities,
which is clearly unphysical. Furthermore, model QGS3.2 predicts
a jump in the optical depth of \ion{Si}{4} relative to \ion{C}{4}
due to the sudden transition in its UVB at $z=3.2$. This is
disfavored by the observations (Aguirre et al. 2004). In
contrast, UVB model QG appears compatible with all the observations 
so far and also gives physical results regarding the evolution of
metallicity with time and density (Schaye et al. 2003;
Aguirre et al. 2004). Therefore, it seems
unlikely that this model could be seriously in error 
(Aguirre et al. 2004). The results on the IGM inventory for UVB 
model QG (see Table 1) will be used in the subsequent presentation.
As shown by QW05, VMS contributions are required to match
[Si/C]$_{\rm IGM}$ for this model and
both [Si/C]$_{\rm IGM}$ and [C/O]$_{\rm IGM}$
for this model can be accounted for by a blend of VMS and SN II 
contributions. Insofar as [Si/C]$_{\rm IGM}$ is approximately at
or above the value given by this model, these results appear robust.
The high value of [Si/C]$_{\rm IGM}$ was not discussed 
in the recent critique of VMS models by Tumlinson,
Venkatesan, \& Shull (2004) or in another study by
Daigne et al. (2004). In particular, Daigne et al. (2004) 
concluded that ``there is
no evidence, nor any need, for a hypothesized
primordial population of very massive stars in order
to account for the chemical abundances of extremely
metal-poor halo stars or of the intergalactic medium.''
We consider that further improvements in measurements and analyses
of the IGM abundances, as well as further theoretical studies of
the conditions under which stars of extremely low metallicity
can form, will clarify the choice between the metal
enrichment scenarios with and without VMSs.

Based on the high value of [Si/C]$_{\rm IGM}$ for UVB model 
QG, we use the VMS model of HW02 and the low-metallicity
SN II model of Woosley \& Weaver (1995, hereafter WW95)
to quantify VMS contributions to the IGM.
The VMSs relevant for metal enrichment are in the mass range 
$M_{\rm VMS}\approx 140$--$260\,M_\odot$ and produce
PI-SNe (HW02). A Salpeter IMF is assumed for both VMSs and SNe II 
and the corresponding yield ratios are given
in Table 1 (models 2A and 6 in Table 2 of QW05). 
Using these yield ratios, we can find mixtures of VMS and SN II
contributions that can account for [Si/C]$_{\rm IGM}$ and 
[C/O]$_{\rm IGM}$ simultaneously. As both [C/O]$_{\rm VMS}=-0.57$ 
and [C/O]$_{\rm SN\ II}=-0.42$ are within 0.1 dex of 
[C/O]$_{\rm IGM}=-0.50$, the appropriate mixture is determined 
by matching [Si/C]$_{\rm IGM}$. The required fraction 
$f_{\rm Si}^{\rm VMS}$ of Si contributed by VMSs can be 
calculated from
\begin{equation}
\left(\frac{\rm C}{\rm Si}\right)_{\rm IGM}=
\left(\frac{\rm C}{\rm Si}\right)_{\rm VMS}
f_{\rm Si}^{\rm VMS}+\left(\frac{\rm C}{\rm Si}\right)_{\rm SN\ II}
(1-f_{\rm Si}^{\rm VMS}).
\end{equation}
With [Si/C]$_{\rm VMS}=1.18$ and [Si/C]$_{\rm SN\ II}=0.42$,
$f_{\rm Si}^{\rm VMS}=0.63$ is required to match 
[Si/C]$_{\rm IGM}=0.74$ exactly and $f_{\rm Si}^{\rm VMS}=0.5$
gives [Si/C]~$=0.65$. Therefore, VMSs must be the dominant 
source of Si. This result does not depend on the specific
stellar models adopted here. In fact, as other SN II models
have lower values of [Si/C]$_{\rm SN\ II}$ (see Table 2 of 
QW05), using them only strengthens the above result.
We conclude that a minimum value of 
$f_{\rm Si}^{\rm VMS}\sim 0.5$ is required to account for the 
observed elemental ratios in the IGM. For the stellar models 
adopted here, $f_{\rm Si}^{\rm VMS}\sim 0.5$ corresponds to 
$f_{\rm C}^{\rm VMS}\sim 0.15$ and 
$f_{\rm O}^{\rm VMS}\sim 0.19$, so SNe II are the dominant 
source of C and O. This conclusion can be 
generalized to all stellar models as well.

\section{Metal Enrichment of the IGM by VMSs}

We now discuss metal enrichment of the IGM by VMSs.
The formation of these stars is supposed to precede that of
lower-mass stars and occur in halos
where big bang debris can cool by H$_2$ molecules 
(see Abel et al. 2002; Bromm \& Larson 2004 for reviews). 
This requires a minimum halo mass of $\sim 10^5\,M_\odot$. 
The efficiency of VMS formation in such low-mass
halos determines how early VMSs could provide substantial 
enrichment to the IGM. For a simple estimate, we consider 
that all halos with mass $M>M_{\rm min}$ are formed from 
``building-block'' halos with mass $M_{\rm min}$. 
We assume that a VMS would have formed in every 
``building-block'' halo and ejected all its nucleosynthetic 
products into the IGM. Then at redshift $z$, 
the abundance of element E in the IGM would be
\begin{equation}
Z_{\rm E}^{\rm IGM}\equiv\frac{{\rm (E/H)}_{\rm IGM}}{{\rm (E/H)}_\odot}
\approx\frac{\langle Y_{\rm E}^{\rm VMS}\rangle 
F(M>M_{\rm min}|z)}{X_{\rm E}^\odot f_b M_{\rm min}},
\end{equation}
where $\langle Y_{\rm E}^{\rm VMS}\rangle$ is the average 
mass yield of E per VMS event, $F(M>M_{\rm min}|z)$ is 
the fraction of all matter residing in halos with $M>M_{\rm min}$ 
at redshift $z$, $X_{\rm E}^\odot$ is the solar mass fraction 
of E, and $f_b$ is the baryonic fraction of the halo mass.
Consider Si as an example. Using
$\langle Y_{\rm Si}^{\rm VMS}\rangle=16.6\,M_\odot$ (HW02), 
$X_{\rm Si}^\odot=7.1\times 10^{-4}$ 
(Anders \& Grevesse 1989), $f_b=0.15$, and
$M_{\rm min}=10^5\,M_\odot$, we obtain
$Z_{\rm Si}^{\rm IGM}\approx 1.6F(M>10^5\,M_\odot|z)$.
The function $F(M>10^5\,M_\odot|z)$ can be estimated using the 
Press-Schechter formalism (Press \& Schechter 1974) and 
ranges from $\sim 0.3$\% to 3\% for $z\sim 24$ to 17. Thus,
if VMSs are to provide 50\% of the measured
inventory of [Si/H]$_{\rm IGM}=\log Z_{\rm Si}^{\rm IGM}=-2.0$
(Aguirre et al. 2004), this requires that a VMS form in
every ``building-block'' halo by $z\sim 24$ or more plausibly,
in $\sim 10$\% of such halos by $z\sim 17$.

For a formal and self-consistent model of the chemical 
evolution of the IGM 
using the rate of occurrence of a VMS event in a halo, 
we consider a large reference region of the
universe and treat it as a closed homogeneous system (QW05). The
evolution of (E/H)$_{\rm IGM}$ in the system as a function of
time $t$ is governed by
\begin{equation}
\frac{d{\rm (E/H)}_{\rm IGM}}{dt}=\frac{Q_{\rm E}}{{\rm (H)}_{\rm IGM}},
\label{dehdt}
\end{equation}
where $Q_{\rm E}$ is the net rate for ejection of E atoms into
the IGM and (H)$_{\rm IGM}$ is the number of H atoms in the IGM.
The rate $Q_{\rm E}$ depends on the rate of occurrence of a VMS 
event and the efficiency of gas expulsion by the VMS explosion.
We consider that a VMS may form in a halo 
when the virial temperature $T_{\rm vir}$ of the gas reaches a 
minimum value of $T_{\rm vir,0}=300$~K for H$_2$ cooling.
For $z\gg 1$,
\begin{equation}
T_{\rm vir}\approx 211\left(\frac{\mu}{1.22}\right)
\left(\frac{M}{10^5\,M_\odot}\right)^{2/3}
\left(\frac{1+z}{10}\right)\ {\rm K}, 
\label{tvir}
\end{equation}
where $\mu$ is the mean atomic weight and $\mu=1.22$ or 0.6 for 
a neutral or ionized gas, respectively, with a primordial 
composition of H and He. We assume that 
all the gas would be expelled 
from the halo by the VMS explosion if the gravitational binding 
energy of the gas $E_{\rm bi,gas}$ is less than the VMS explosion 
energy $E_{\rm exp}$ (e.g., Bromm, Yoshida, \& Hernquist 2003) and 
retained for $E_{\rm bi,gas}\geq E_{\rm exp}$. For $z\gg 1$,
\begin{equation}
E_{\rm bi,gas}\approx 4.31\times 10^{47}
\left(\frac{M}{10^5\,M_\odot}\right)^{5/3}
\left(\frac{1+z}{10}\right)\ {\rm erg}. 
\label{ebgas}
\end{equation}
For a given $z$, a halo associated with an $n_0\ \sigma$ density 
fluctuation (an $n_0\ \sigma$ halo) would have the mass $M_0$ 
corresponding to $T_{\rm vir}=T_{\rm vir,0}$ for the onset of VMS
formation, while an $n_{\rm bi}\ \sigma$ halo would have the mass 
$M_{\rm bi}$ corresponding to $E_{\rm bi,gas}=E_{\rm exp}$ for the
onset of gas retention. Then we have $Q_{\rm E}=
\sum_{n=n_0}^{n_{\rm bi}}N_{n\ \sigma}P_{{\rm E},n\ \sigma}$,
where $N_{n\ \sigma}$ is the total number of halos with a specific 
$n$ value in the system and $P_{{\rm E},n\ \sigma}$ is the number 
production rate of E in a single $n\ \sigma$ halo.

The rate $P_{{\rm E},n\ \sigma}$ is related to the rate of 
occurrence of a VMS event $R_{n\ \sigma}$ in an $n\ \sigma$ halo.
We assume that these rates are proportional to the 
number (H)$_{n\ \sigma}$ of H atoms in the gas of the 
halo and write
\begin{eqnarray}
P_{{\rm E},n\ \sigma}&=&\langle y_{\rm E}^{\rm VMS}\rangle R_{n\ \sigma}
=\Lambda_{\rm E}^{\rm VMS}{\rm (E/H)}_\odot({\rm H})_{n\ \sigma},
\label{pe}\\
R_{n\ \sigma}&\approx&
\Lambda_{\rm E}^{\rm VMS}X_{\rm E}^\odot\frac{f_bM_{n\ \sigma}}
{\langle Y_{\rm E}^{\rm VMS}\rangle},
\label{rn}
\end{eqnarray}
where $\langle y_{\rm E}^{\rm VMS}\rangle$ is the average number 
yield of E per VMS event corresponding to the mass yield 
$\langle Y_{\rm E}^{\rm VMS}\rangle$ and 
$\Lambda_{\rm E}^{\rm VMS}$ is a rate constant. 
The above prescription of
$P_{{\rm E},n\ \sigma}\propto R_{n\ \sigma}\propto
({\rm H})_{n\ \sigma}$ is meant to imply that it is more probable 
for a VMS to form in a larger halo, and we will discuss the limit
on the occurrence of a VMS event in a typical halo below.
Substituting $Q_{\rm E}$ in equation~(\ref{dehdt}) and noting that 
for $z\gg 1$ the majority of the H atoms of the system resides 
in the IGM and the majority of the H atoms in a halo resides in
the gas, we obtain
\begin{eqnarray}
\frac{dZ^{\rm IGM}_{\rm E}}{dt}&=&\Lambda_{\rm E}^{\rm VMS}
\sum_{n=n_0}^{n_{\rm bi}}N_{n\ \sigma}({\rm H})_{n\ \sigma}/{\rm (H)}_{\rm IGM}
\approx\Lambda_{\rm E}^{\rm VMS}F(M_0<M<M_{\rm bi}|z),\label{dzdt}\\
Z_{\rm E}^{\rm IGM}(t)&\approx&\Lambda_{\rm E}^{\rm VMS}\int^{t(z)}_0 
F(M_0<M<M_{\rm bi}|z')dt',
\label{zigm}
\end{eqnarray}
where 
\begin{equation}
F(M_0<M<M_{\rm bi}|z)=\sqrt{\frac{2}{\pi}}\int_{n_0}^{n_{\rm bi}}
\exp\left(-\frac{x^2}{2}\right)dx
\label{f0bi}
\end{equation}
is the fraction of all matter residing in $n\ \sigma$ halos with 
$n_0<n<n_{\rm bi}$ as given by the Press-Schechter formalism 
and $t(z)=0.538[10/(1 + z)]^{3/2}$~Gyr 
for the adopted cosmology 
(see discussion in Barkana \& Loeb 2001).

The evolution of $Z_{\rm E}^{\rm IGM}$ is thus only dependent 
on $\Lambda_{\rm E}^{\rm VMS}$ and $F(M_0<M<M_{\rm bi}|z)$. As 
$Z_{\rm E}^{\rm IGM}$ scales with $\Lambda_{\rm E}^{\rm VMS}$, 
we will use $\Lambda_{\rm E}^{\rm VMS}=0.1$~Gyr$^{-1}$ as a 
reference value (for elements produced by SNe II only, the average
Galactic value of $\Lambda_{\rm E}^{\rm SN\ II}$ is 
$\sim 0.1$~Gyr$^{-1}$ 
over a period of $\sim 10$~Gyr prior to solar system formation;
QW05). To determine $F(M_0<M<M_{\rm bi}|z)$, we use 
$T_{\rm vir,0}=300$~K ($\mu=1.22$) to find $M_0$ and a 
typical VMS explosion energy of $E_{\rm exp}=4\times 10^{52}$~erg 
(HW02) to find $M_{\rm bi}$ for which $E_{\rm bi,gas}=E_{\rm exp}$.
The relevant range of $n\ \sigma$ halos with
$M_0<M<M_{\rm bi}$ ($n_0<n<n_{\rm bi}$) corresponds to
$5.5\times 10^4\,M_\odot<M<6.1\times 10^7\,M_\odot$
($2.3<n<3.8$) at $z=20$ and
$8.3\times 10^4\,M_\odot<M<7.2\times 10^7\,M_\odot$
($1.8<n<2.9$) at $z=15$. Although the range $M_0<M<M_{\rm bi}$ is
rather wide, the integral for $F(M_0<M<M_{\rm bi}|z)$ is
dominated by the contributions from $n\ \sigma$ halos with
$M\sim M_0$ ($n\sim n_0$; see eq.~[\ref{f0bi}]). Thus,
a typical halo relevant for metal-enrichment of the IGM has 
$M\sim 10^5\,M_\odot$ for $z\sim 15$--20 and an exact treatment of
halos of much higher masses, which requires a detailed description of
their merger history, is not crucial to our model. 
[In particular, $F(M_0<M<M_{\rm bi}|z)$ 
is not sensitive to $E_{\rm exp}$
for $E_{\rm exp}\gtrsim 10^{51}$~erg
as halos with $E_{\rm bi,gas}\gtrsim 10^{51}$~erg 
are extremely rare for $z\gg 1$]. The evolution of 
[E/H]$_{\rm IGM}=\log Z_{\rm E}^{\rm IGM}$ for 
$\Lambda_{\rm E}^{\rm VMS}=0.1$~Gyr$^{-1}$ is shown as the
dot-dashed curve in Figure~1. This gives
[E/H]$_{\rm IGM}=-3.46$ at $z=15$. Thus, to provide 50\% of the 
measured inventory of [Si/H]$_{\rm IGM}=-2.0$ by $z=15$ 
(point A) requires
$\Lambda_{\rm Si}^{\rm VMS}=1.4$~Gyr$^{-1}$, which corresponds 
to $R_{n\ \sigma}t_{15}\sim 0.2$ VMS event over 
$t_{15}\equiv t(z=15)=0.27$~Gyr
for a halo with $M_{n\ \sigma}\sim 10^5\,M_\odot$ and
$R_{n\ \sigma}\sim 0.9$~Gyr$^{-1}$ (eq.~[\ref{rn}]).
In Figure~2 we show the values of $\Lambda_{\rm Si}^{\rm VMS}$ 
necessary to achieve [Si/H]~$=-2.3$ (solid curve) or $-2.0$ 
(dashed curve) by a given $z$ as calculated from 
equation~(\ref{zigm}). The dot-dashed curve indicates
formation of $R_{n\ \sigma}t(z)\sim 1$ VMS by time $t(z)$ in 
a halo of mass $M_{n\ \sigma}\sim 10^5\,M_\odot$ and is taken 
as a bound. It is evident 
that a major part ($\gtrsim 50$\%) of the Si in the IGM, which
must come from VMSs, can only be provided for $z$ substantially 
below 19 and with plausible production rates, for $z<16$.
We consider $\Lambda_{\rm Si}^{\rm VMS}=1.4$~Gyr$^{-1}$ 
a reasonable rate and show the corresponding evolution of 
[Si/H]$_{\rm IGM}$ for $z>15$ as the solid curve in Figure~1. 
The evolution of [C/H]$_{\rm IGM}$ and [O/H]$_{\rm IGM}$ is fixed
by the yield ratios of the VMS model (see Table~1) and shown 
as the dot-dashed and short-dashed curves, respectively.

\section{Termination of VMS Contributions}

Figure~1 shows that for a reasonable formation efficiency,
VMSs would provide [C/H]~$=-3.5$ and [O/H]~$=-2.9$ to the IGM 
by $z=15$. This is consistent with the proposal by
Bromm \& Loeb (2003) that VMSs can no longer 
be formed when [C/H]~$=-3.5\pm 0.1$ and [O/H]~$=-3.05\pm 0.2$ 
are reached in the IGM. They showed that for these
metallicities, gas clouds initially cooled by H$_2$ molecules
could continue to cool by C and O atoms and fragment into 
smaller clumps. However, the termination of VMS formation
is a complex matter. The soft UV radiation from VMSs 
would dissociate H$_2$ molecules in low-mass halos,
thus suppressing later VMS formation. The history of H$_2$
dissociation is not well known but a theoretical study by
Ciardi et al. (2000) suggested that 
universal dissociation has not yet occurred for $z\sim 20$.
The UVB produced by VMSs may also 
play an important role in reionization, which could have 
taken place at $z\sim 17$ (Kogut et al. 2003). As shown by 
Oh et al. (2001), there should be sufficient hard UV photons
emitted from VMSs to reionize the universe if these stars
were to provide [Si/H]$_{\rm IGM}\sim -2.3$ 
(see Tumlinson et al. 2004; Daigne et al. 2004 for
a different view).
The above estimates for the onset of C and O cooling,
universal H$_2$ dissociation, and reionization appear to 
roughly coincide. Clearly, there must be a transition region
in which ongoing VMS formation increases the soft UVB for
H$_2$ dissociation and begins to suppress further
VMS formation, thus resulting in a global 
decrease in $\Lambda_{\rm E}^{\rm VMS}$. There will also be 
an accompanying increase in the hard UVB that begins to
initiate reionization and raise the IGM temperature.
The balance between these processes is not well understood
and cannot be addressed here. It is possible that both 
effects are operating to end the regime of H$_2$ cooling
at roughly similar $z$ to reionization. 
Combining the consideration of the increasing UVB with the 
metallicity condition for termination of VMS 
formation proposed by Bromm \& Loeb (2003),
we will assume that VMS contributions to the IGM will cease 
at $z\sim 15$ and transition to formation of lower-mass stars 
will occur universally.

\section{Enrichment of the IGM by Galactic Outflows}

For simplicity we assume that VMS production stops sharply
at $z=15$ and take VMS contributions to the IGM as shown in 
Figure~1. Then the bulk of the C and O and the remainder of 
the Si must be provided by SNe II at $15>z\gtrsim 4$. 
The formalism resulting in equation~(\ref{dzdt}) can be modified
to treat the chemical evolution of the IGM in this regime,
where formation of regular stars including SN II 
progenitors requires $T_{\rm vir,0}=10^4$~K 
($\mu=0.6$) for cooling by atomic species. 
This corresponds to halos with $M\sim 10^8\,M_\odot$ and
$E_{\rm bi,gas}$ far exceeding the typical
SN II explosion energy of $\sim 10^{51}$~erg.
Thus only a fraction $\epsilon$ of the debris from an SN II
will be ejected into the IGM.
While the dependence of $\epsilon$ on the halo mass may be 
complicated, we will treat it as a constant for halos below
some cut-off mass $M_1$. Our goal is to estimate how efficient 
SN II-driven galactic outflows must be in order to provide a 
substantial part of the IGM inventory at $15>z\gtrsim 4$. 
Equation~(\ref{dzdt}) may be rewritten in this regime as
\begin{eqnarray}
\frac{dZ_{\rm E}^{\rm IGM}}{dt}&\approx&\epsilon\Lambda_{\rm E}^{\rm SN\ II} 
F(M_0 < M < M_1|z),\\
Z_{\rm E}^{\rm IGM}(t)&\approx& Z_{\rm E}^{\rm IGM}(t_{15}) +
\epsilon\Lambda_{\rm E}^{\rm SN\ II}\int_{t_{15}}^{t(z)}F(M_0 < M < M_1|z')dt',
\end{eqnarray}
where $Z_{\rm E}^{\rm IGM}(t_{15})$ is the IGM
inventory of E at $z=15$ resulting from just VMS production.

Note that $Z_{\rm E}^{\rm IGM}(t)-Z_{\rm E}^{\rm IGM}(t_{15})$ 
scales with $\epsilon\Lambda_{\rm E}^{\rm SN\ II}$. 
For convenience we use 
$\epsilon\Lambda_{\rm E}^{\rm SN\ II}=0.01$~Gyr$^{-1}$ as a 
reference value. This corresponds to an 
outflow efficiency of $\epsilon=0.1$ for
a Galactic SN II production rate of
$\Lambda_{\rm E}^{\rm SN\ II}=0.1$~Gyr$^{-1}$.
Using $T_{\rm vir,0}=10^4$~K and $M_1=10^{10}\,M_\odot$, we show
the contributions from just galactic outflows as the 
long-dashed curve in Figure~1. The value of $M_1=10^{10}\,M_\odot$
is consistent with observations of outflows from Lyman break and
dwarf galaxies (e.g., Pettini et al. 2001; Martin et al. 2002).
The results will not change much if $M_1$ is varied from this
value by a factor of several either way (QW05). As can be seen
from Figure~1, galactic outflows 
can only provide [E/H]$_{\rm IGM}=-2.92$ by $z=4$
for $\epsilon\Lambda_{\rm E}^{\rm SN\ II}=0.01$~Gyr$^{-1}$. 
To provide [Si/H]~$=-2.3$ (50\% of the IGM inventory) by $z=4$ 
would require
$\epsilon\Lambda_{\rm Si}^{\rm SN\ II}=0.042$~Gyr$^{-1}$.
This corresponds to a rather high outflow efficiency of 
$\epsilon=0.42$ for intermediate-mass 
($\sim 10^8$--$10^{10}\,M_\odot$) halos 
with a Galactic production rate. If galactic outflows are
to provide [Si/H]~$=-2.3$ by $z=3$, then
$\epsilon\Lambda_{\rm Si}^{\rm SN\ II}=0.024$~Gyr$^{-1}$ is
required and $\epsilon=0.24$ is sufficient
for a Galactic production rate.

Figure~1 shows an example of a self-consistent model for the
chemical evolution of the IGM with contributions from VMSs for
$z>15$ and galactic outflows for $z<15$. The overall evolution 
of [Si/H]$_{\rm IGM}$, [O/H]$_{\rm IGM}$, and [C/H]$_{\rm IGM}$
is shown assuming that VMSs provided 50\% of the Si inventory 
by $z=15$ (point A) and galactic outflows the other 
half by $z=4$ (point B) and using the yield ratios of the 
sources (see Table~1). It can be seen that galactic outflows 
cannot provide significant contributions until $z\sim 6$. 
Thus, the model implies that there should be very little 
change in the net IGM inventory between $z=15$ and $z\sim 6$.
The respective contributions to C, O, Si, and Fe from VMSs
and galactic outflows associated with SNe II are given in
Table 1.

\section{Metals in the IGM and Inventory of Massive Black Holes}

In considering metal enrichment of the IGM at high $z$ we have
focused on VMSs with $M_{\rm VMS}\approx 140$--$260\,M_\odot$
in low-mass halos using the yields of HW02. These workers showed
that this narrow range of stellar masses produce 
PI-SNe, which would efficiently eject the associated nucleosynthetic
products into the IGM. They also showed that all 
VMSs with $M_{\rm VMS}>260\,M_\odot$ would form black
holes and therefore not contribute any metals. Thus, the first 
stellar sources contributing metals are restricted to PI-SNe. However, 
if one considers any plausible IMF, the occurrence of PI-SNe implies 
the production of more massive stars that will thus give massive black 
holes (MBHs). It follows that the IGM inventory of metals that is 
attributable to PI-SNe must be coupled with the production 
of MBHs. This then relates the metal (particularly Si) content of the IGM 
to the inventory of MBHs at early epochs. 

Before discussing the coupling between the amount of metals in the IGM
and the total mass of MBHs produced in epochs prior to galaxy formation,
we note the work by Madau \& Rees (2001), who
estimated the inventory of MBHs based on considerations other than metal
enrichment. Motivated by simulations of
Bromm, Coppi, \& Larson (1999) and Abel, Bryan, \& Norman (2000), which 
suggested that the first stars may have been very massive, Madau \& Rees 
(2001) recognized the significance of possible early production of MBHs 
in an episode of pregalactic star formation using earlier stellar models
(Bond, Arnett, \& Carr 1984; Fryer, Woosley \& Heger 2001), which showed 
that VMSs could produce MBHs. In 
particular, they found that if one MBH of mass $M_{\rm MBH}\gtrsim
150\,M_\odot$ formed in each $3\ \sigma$ halo 
(with $3\times 10^5h^{-1}\,M_\odot$ in dark and baryonic matter) 
collapsing at $z\approx 20$, then a fraction
$f_{\rm MBH}\gtrsim 8\times 10^{-5}h^3$ of all baryonic matter would be 
in MBHs, where $h$ is the Hubble parameter in units of 
100~km~s$^{-1}$~Mpc$^{-1}$ and we take $h=0.7$ throughout this paper. 
This estimate is comparable to the total mass fraction of the central
supermassive black holes (SMBHs) found in most nearby galaxies
(see eq.~[\ref{fsmbh}] below).

In the earlier sections of the present paper, we have inferred that
VMSs with $M_{\rm VMS}\approx 140$--$260\,M_\odot$ provided  
[Si/H]$_{\rm IGM} = -2.3$ ($\sim 50$\% of the Si in the IGM) and we 
have derived the rate of VMS formation required to achieve this. 
We now explore the implications of this metal
production for the inventory of MBHs that would be produced in low-mass 
halos at the same epoch and
show that it is in accord with the possibility considered 
by Madau \& Rees (2001).
For a quantitative estimate, we consider a large reference region 
of the IGM with a total mass $M_{\rm IGM}$ and use a
Salpeter IMF for VMSs. To enrich this IGM with 
[Si/H]$_{\rm IGM}=\log Z_{\rm Si}^{\rm IGM}$,
the required number $N_{\rm VMS,Si}$ of VMSs with 
$M_{\rm VMS}\approx 140$--$260\,M_\odot$ is
\begin{equation}
N_{\rm VMS,Si}\approx\frac{Z_{\rm Si}^{\rm IGM}
X_{\rm Si}^\odot M_{\rm IGM}}
{\langle Y_{\rm Si}^{\rm VMS}\rangle}.
\end{equation}
As the total baryonic mass of a hosting halo is
$\sim 10^4\,M_\odot$, we take the upper mass limit for VMSs
to be $2000\,M_\odot$. We have checked that our results are not
sensitive to this assumed upper mass limit.
The number $N_{\rm MBH}$ of MBHs 
resulting from VMSs with $M_{\rm VMS}=260$--$2000\,M_\odot$ is
\begin{equation}
N_{\rm MBH}\approx\frac{\int_{260}^{2000}m^{-2.35}dm}
{\int_{140}^{260}m^{-2.35}dm}N_{\rm VMS,Si}
\approx 0.72N_{\rm VMS,Si},
\label{nmbh}
\end{equation}
where $m$ is the VMS mass in units of $M_\odot$. The average
mass of these MBHs is
\begin{equation}
\langle M_{\rm MBH}\rangle\approx\alpha
\frac{\int_{260}^{2000}m^{-1.35}dm}{\int_{260}^{2000}m^{-2.35}dm}M_\odot
\approx 547\alpha M_\odot,
\label{mmbh}
\end{equation}
where $\alpha\sim 0.5$--1 is the average mass fraction of a VMS 
progenitor in the relevant mass range
that ends up in the MBH (Fryer et al. 2001). 
The fractional contribution of MBHs to the baryonic mass of the 
universe can be estimated as
\begin{eqnarray}
f_{\rm MBH}&\approx&\frac{N_{\rm MBH}\langle 
M_{\rm MBH}\rangle}{M_{\rm IGM}}
\approx Z_{\rm Si}^{\rm IGM}X_{\rm Si}^\odot 
\left(\frac{N_{\rm MBH} }{N_{\rm VMS,Si}}\right)
\left(\frac{\langle M_{\rm MBH}\rangle}
{\langle Y_{\rm Si}^{\rm VMS}\rangle} \right)\\
&\approx&8.4\times 10^{-5}\alpha\left(\frac{Z_{\rm Si}^{\rm IGM}}
{5\times 10^{-3}}\right).
\label{fmbh}
\end{eqnarray}

The above estimate\footnote{As stated above, this result 
assumes a Salpeter IMF ($\propto m^{-2.35}$) for VMSs. For a steeper 
IMF of the form $\propto m^{-3}$, the numerical coefficient in
eq.~(\ref{fmbh}) for $f_{\rm MBH}$ is reduced by a factor
of $\approx 2$.} of $f_{\rm MBH}$ may be compared with the fraction
$f_{\rm SMBH}$ of all baryonic matter contributed by SMBHs in the
present universe,
\begin{equation}
f_{\rm SMBH}=\frac{\rho_{\rm SMBH}}{\Omega_b\rho_{\rm cri}}= 
7.5^{+3.1}_{-2.3}\times 10^{-5},
\label{fsmbh}
\end{equation}
where $\rho_{\rm SMBH}=4.6^{+1.9}_{-1.4}\times 
10^5\,M_\odot$~Mpc$^{-3}$ is the present mass density of SMBHs
obtained from galactic observations
(see the recent analyses by Marconi et al. 2004), 
$\rho_{\rm cri} =1.36\times 10^{11}\,M_\odot$~Mpc$^{-3}$ is 
the present critical density, and $\Omega_b=0.045$ 
is the fractional contribution of baryons to $\rho_{\rm cri}$.
It can be seen that the mass fraction of MBHs calculated from
the Si abundance in the IGM of the early universe (eq.~[\ref{fmbh}]) 
and that of SMBHs determined for the present epoch 
by observations (eq.~[\ref{fsmbh}]) 
are in remarkably good accord. The possibility considered by 
Madau \& Rees (2001) is now shown to be supported by the required VMS 
contribution to produce the Si in the IGM. We note that 
Schneider et al. (2002) derived a wide range of formation efficiency 
for VMSs with $M_{\rm VMS}\approx 140$--$260\,M_\odot$ using
$f_{\rm MBH}\leq f_{\rm SMBH}$ and other constraints. Their results 
can be easily accommodated by the rate of VMS formation derived here. 
We also note that accreting MBHs may contribute to the soft X-ray 
background (SXRB). Based on the unaccounted SXRB flux, 
Salvaterra, Haardt, \& Ferrara (2005) estimated that $f_{\rm MBH}$ 
cannot exceed $\sim 10^{-4}$. This is consistent with our result. 
During the epoch of VMS formation at $z\gtrsim 15$, the rate of 
occurrence of an MBH-producing VMS event is
$(N_{\rm MBH}/N_{\rm VMS,Si})R_{n\ \sigma}\sim 0.6$~Gyr$^{-1}$
for a halo of $\sim 10^5\,M_\odot$ (eq.~[7] with 
$\Lambda_{\rm Si}^{\rm VMS}=1.4$~Gyr$^{-1}$ and eq.~[\ref{nmbh}]).
This rate can be used to calculate the contribution from 
MBHs to the SXRB.

The remarkable accord between the MBH inventory calculated from
the Si content of the early IGM (eq.~[\ref{fmbh}]) and the present
global mass budget of SMBHs obtained from galactic observations 
(eq.~\ref{fsmbh}]) cannot be easily dismissed as an accident. It is
expected that the early-formed MBHs would cluster near 
galactic centers (Madau \& Rees 2001). As a fraction 
$\sim (4$--$8)\times 10^{-5}$ of all baryonic matter is in MBHs 
(eq.~[\ref{fmbh}] with $\alpha\sim 0.5$--1), a typical galaxy 
with a baryonic mass of $\sim 10^{11}\,M_\odot$ would have
a total mass of $\sim (4$--$8)\times 10^6\,M_\odot$ in MBHs. 
This is close to the mass of $(3.7\pm 0.4)\times 10^6\,M_\odot$ 
for the SMBH at the Galactic center (e.g., Ghez 2004). 
It is possible that the central SMBHs in typical
galaxies of the present universe are formed by simply merging
the MBH seeds without much gas accretion. This would explain
why the total mass of SMBHs observed at the present epoch corresponds 
approximately (within a factor of $\sim 2$) to that of the MBH 
seeds. 

However, there appears to be an obvious conflict. It is widely 
accepted that quasars are powered by gas accretion onto SMBHs. 
The total gas mass accreted by all optical quasars is inferred to be 
also comparable to that of the SMBHs observed at the present epoch
(e.g., Yu \& Tremaine 2002; Marconi et al. 2004). This is supported 
by the simulations of Hopkins et al. (2005), who proposed a unified 
model for producing SMBHs, quasars, and galaxy spheroids. In their 
model, SMBHs were formed predominantly by gas accretion. 
Both the global mass budget and the relative distribution of SMBHs 
over the mass range $M_{\rm SMBH}\sim 10^6$--$10^{10}\,M_\odot$ given 
by their model are in agreement with those derived by Marconi et al. 
(2004) from galactic observations. It appears that the model of
Hopkins et al. (2005) is rather satisfactory. However, within 
the uncertainties of their model, it is still plausible to consider 
that the more common SMBHs of $<10^8\,M_\odot$ might represent 
aggregates of the MBH seeds without much gas accretion
but large amounts of gas accretion must 
occur to explain the rarer SMBHs of $>10^8\,M_\odot$. Thus, in
accounting for the approximate agreement between the global SMBH 
mass budget determined from galactic and quasar observations and
the MBH inventory, there appears to be a ``rule'' that only the 
rarer higher-mass SMBHs can efficiently accrete gas
but the more common 
lower-mass ones cannot. This matter clearly requires attention. 
In the model of
Hopkins et al. (2005), mergers of halos enhance the infall of gas
toward the SMBHs. However, as mergers must have also occurred to
form galaxies hosting SMBHs of $<10^8\,M_\odot$, which constitute 
$\approx 50$\% of the global SMBH mass budget (Marconi et al. 2004), 
there seem to be some other processes that inhibit gas accretion onto 
such lower-mass SMBHs.

The complex problem of forming SMBHs from MBH seeds was studied by 
Volonteri, Haardt, \& Madau (2003). However, the total mass of 
MBHs in their scenarios is only a small fraction (typically 
$\sim 10^{-3}$) of 
that of the SMBHs at the present epoch as cited above. We 
urge that the problem of forming SMBHs from MBH seeds be revisited 
with the initial conditions presented here. For a formation rate
of $\sim 0.6$~Gyr$^{-1}$ in each halo of $\sim 10^5\,M_\odot$, an MBH 
with an average mass 
$\langle M_{\rm MBH}\rangle\sim 270$--$550\,M_\odot$ (eq.~[\ref{mmbh}] 
with $\alpha\sim 0.5$--1) should have been produced in $\sim 16$\% of 
such halos by $z=15$. If the amount of gas accretion onto black holes 
in galactic centers is typically comparable to or smaller than
the total mass of the original inherited MBHs, then there would
not be a serious difficulty in reconciling the global SMBH mass 
budget determined from galactic observations with that from quasar 
observations and with the inventory of MBHs calculated from the VMS 
contribution to the Si in the IGM.

The evolution of an initial distribution of MBH seeds was studied
extensively by Islam, Taylor, \& Silk (2003, 2004) using a 
semi-analytical model to follow the hierarchical merging of halos.
These authors focused on the dynamics of the MBH inventory and
presented a clear analysis of what might be expected from 
assemblage of MBHs without gas accretion. Amongst their conclusions
are: (1) Hierarchical merging of MBH seeds that formed in 
$\sim 3\ \sigma$ halos collapsing at $z\sim 25$ can contribute 
$\gtrsim 10$\% of the present global mass budget of SMBHs.
A central SMBH of $\sim 3\times 10^6\,M_\odot$ in a galaxy like ours
could be the result of seed accumulation without gas accretion. 
However, gas accretion is necessary for higher-mass SMBHs.
(2) For a present-day galaxy of $\sim 10^{10}$--$10^{13}\,M_\odot$
(mostly in dark matter), the total mass of the inherited MBHs in
the galactic halo is comparable to or greater than the mass
contributed by MBHs to the central SMBH. It would be of interest
to see what a new calculation of the type presented by 
Hopkins et al. (2005) would give if the mass and production history
of MBHs as provided here were used. If the majority of MBHs are
stored in galactic halos but not bulges, then we might expect gas
accretion to account for the growth of central SMBHs while the
MBHs in the low-density halos might not accrete. In this case,
it is accidental that the total mass of MBHs calculated from
the VMS contribution to the Si in the IGM is close to the present 
global mass budget of SMBHs.

\section{Conclusions}

A two-stage model for the chemical evolution of the IGM is
proposed considering VMSs in low-mass 
($\sim 10^5\,M_\odot$) halos during early epochs ($z\gtrsim 15$).
This early stage ends when universal H$_2$ dissociation and
reionization occur and is followed by an extended 
quiescent period of
little metal production until sufficient SNe II could occur
in intermediate-mass ($\sim 10^8$--$10^{10}\,M_\odot$) halos to
drive significant outflows. Galactic
outflows contribute mainly at $z\sim 4$--6.
The requirement of early VMS contributions follows from the
high value of [Si/C]~$\sim 0.7$ inferred for the IGM
(Aguirre et al. 2004; see discussion
therein regarding possible uncertainties) and model 
yields of VMSs and SNe II (e.g., HW02; WW95). The bulk
($\gtrsim 50$\%) of the Si must come from VMSs, while
the C and O are predominantly from the later galactic outflows.
The required VMS formation efficiency corresponds to
$\sim 0.2$ VMS per low-mass halo. The requirement on galactic 
outflows implies efficient ($\sim 40$\%) loss of SN II debris
from intermediate-mass halos with a Galactic SN II rate.
Contributions from galactic outflows 
would be diminished as the halo masses and the amount of 
baryonic matter stored in low-mass stars increase. This may
explain the lack of evolution of the IGM inventory for
$z=1.8$--4.1.

The VMS contributions are consistent with the
level of metallicities at which VMS formation would be
suppressed (Bromm \& Loeb 2003). They also correspond to
a sufficient number of UV photons from VMSs to reionize 
the universe (Oh et al. 2001). However, how the increasing 
photon production by VMSs leads to H$_2$ dissociation and 
reionization, which then terminates VMS formation, is a 
complex issue and not addressable here. There is a hint that
all these events occur at similar $z$. It would be a useful
test if H$_2$ dissociation and reionization could be modeled
with the VMS formation rates proposed here.

In producing [Si/H]~$=-2.3$, VMSs also provide [Fe/H]~$=-2.9$
to the IGM (see Table 1). This Fe abundance was considered by
us (e.g., Wasserburg \& Qian 2000;  Qian \& Wasserburg 2002) 
as the onset of regular 
star formation based on observations of low-metallicity
Galactic halo stars. There is a basic issue regarding the
metallicity at which regular stars may be formed.
As C and O atoms provide the important cooling,
Bromm \& Loeb (2003) proposed conditions for regular star 
formation in terms of threshold C and O abundances.
These abundances coincide with the VMS contributions
in our model. In this sense the threshold Fe abundance 
proposed by us is equivalent to the threshold C and O 
abundances. However, the low-mass star HE~0107--5240 cannot
be explained by our scenario. This special star has [C/H] 
well above and [O/H] at or below the threshold but 
[Fe/H]~$=-5.3$ (Christlieb et al. 2004; Bessell, 
Christlieb, \& Gustafsson 2004). Whether some low-mass stars 
such as this may form along with VMSs remains an 
open issue. 

A number of Galactic halo stars are observed to have
$-4\lesssim {\rm [Fe/H]}<-3$. If they formed at
$z\lesssim 4$, this would pose a problem as our model gives
a high value of [Fe/H]$_{\rm IGM}=-2.3$ (see Table 1).
However, the Fe abundances of these stars may be explained
if they formed at $6\lesssim z<15$. The IGM then would have
a net inventory of [Fe/H]~$\sim -2.9$ but the Fe abundances
in different regions would follow a lognormal distribution
with a scatter of $\sim 0.75$~dex (Schaye et al. 2003; 
Simcoe et al. 2004). This leads us to suggest that
some parts of the Galaxy formed from under-enriched regions
of the IGM. 

In conjunction with the required metal production by VMSs
in the narrow mass range $M_{\rm VMS}\approx 140$--$260\,M_\odot$,
there would also be MBHs produced by more massive stars (HW02).
Both of these processes should occur in low-mass halos for
$z\gtrsim 15$. This leads to a relationship between the metal
(particularly Si) content of the early IGM and the inventory of
MBHs ($\langle M_{\rm MBH}\rangle\sim 270$--$550\,M_\odot$) at the 
same epoch. The total mass of MBHs calculated from the VMS contribution
to the Si in the early IGM using a Salpeter IMF
is in remarkable agreement with that of
SMBHs observed at the present epoch.
These early-formed MBHs could cluster near
galactic centers during the later epochs of galaxy formation when
galactic outflows contributed to the IGM. Such a cluster of MBHs
would, if coalesced, form an SMBH of 
$\sim (4$--$8)\times 10^6\,M_\odot$ in a typical galaxy with a 
baryonic mass of $\sim 10^{11}\,M_\odot$. There is a hint of
quasi-conservation of black hole masses, which implies that only the
rarer SMBHs of $>10^8\,M_\odot$ can efficiently accrete gas but the
more common ones of $<10^8\,M_\odot$ cannot. It remains to be
investigated whether a mechanism exists to inhibit gas accretion
onto lower-mass SMBHs.
In conclusion, early ``chemistry'' appears to provide 
considerable insights into aspects of cosmological problems
and offer many intriguing possibilities regarding larger
cosmological issues.

\acknowledgments
We wish to thank Martin Rees for suggesting that we write this
short report separate from any monographic effort. We also would
like to thank the anonymous referee for thorough and incisive
comments that greatly improved the paper.
This work was supported in part by DOE grants DE-FG02-87ER40328
(Y. Z. Q.) and DE-FG03-88ER13851 (G. J. W.),
Caltech Division Contribution 9131(1120).

\clearpage
\begin{figure}
\includegraphics[angle=270,scale=.70]{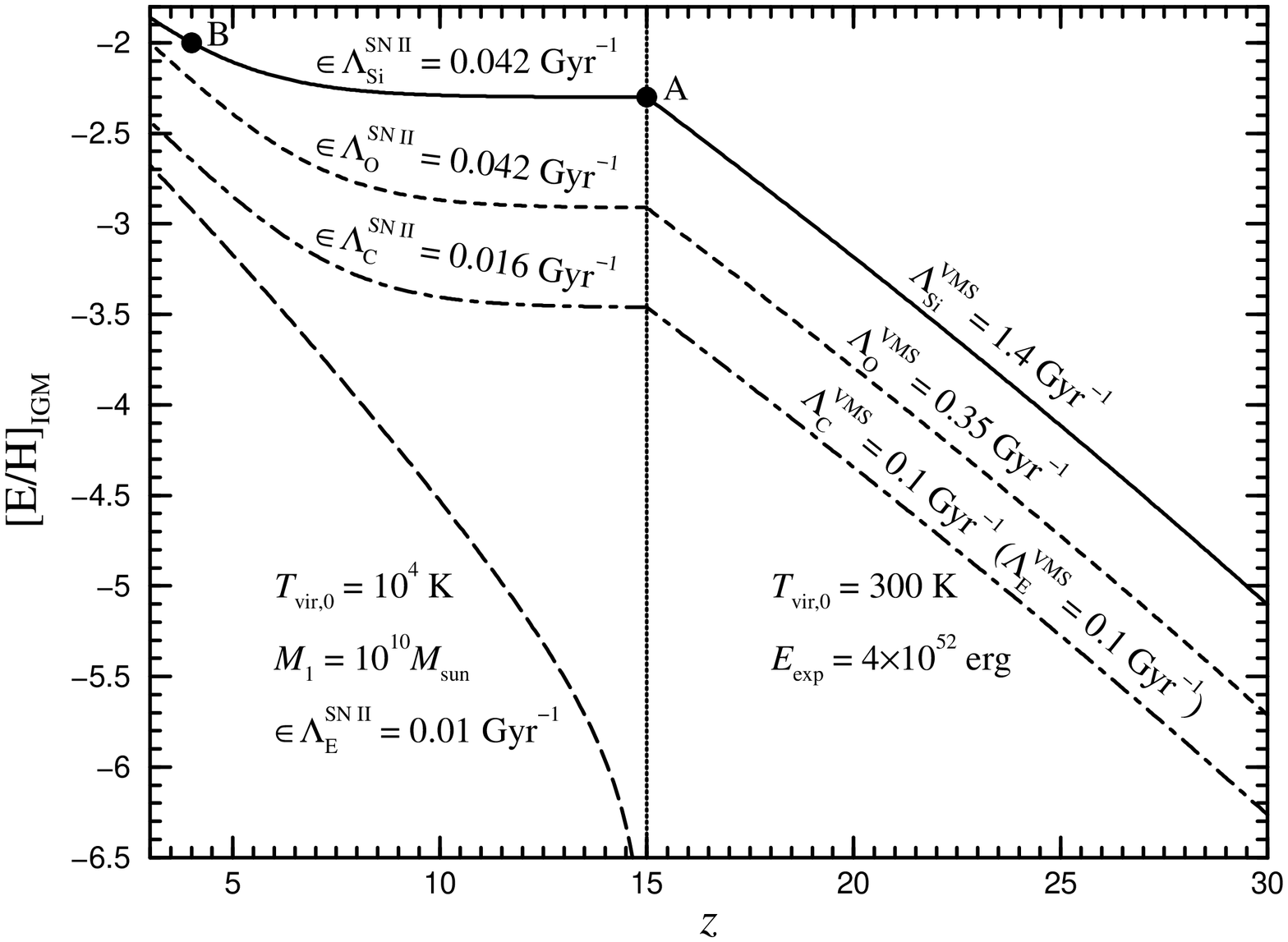}
\caption{Model for evolution of abundances of Si (solid curve),
O (short-dashed curve), and C (dot-dashed curve).
The sources are taken to be VMSs (yields from HW02) in low-mass
halos for $z>15$ and galactic outflows of SN II products
(yields from WW95) from intermediate-mass halos for $z<15$.
The evolution is governed
by the rates $\Lambda_{\rm E}^{\rm VMS}$ and 
$\epsilon\Lambda_{\rm E}^{\rm SN\ II}$. For Si these are chosen to
give [Si/H]~$=-2.3$ at $z=15$ (point A) and the full IGM inventory
of [Si/H]~$=-2.0$ at $z=4$ (point B). The rates for O and C are
fixed by the yields of the sources. The long-dashed curve
represents the evolution resulting from just galactic outflows.}
\end{figure}

\clearpage
\begin{figure}
\epsscale{.80}
\plotone{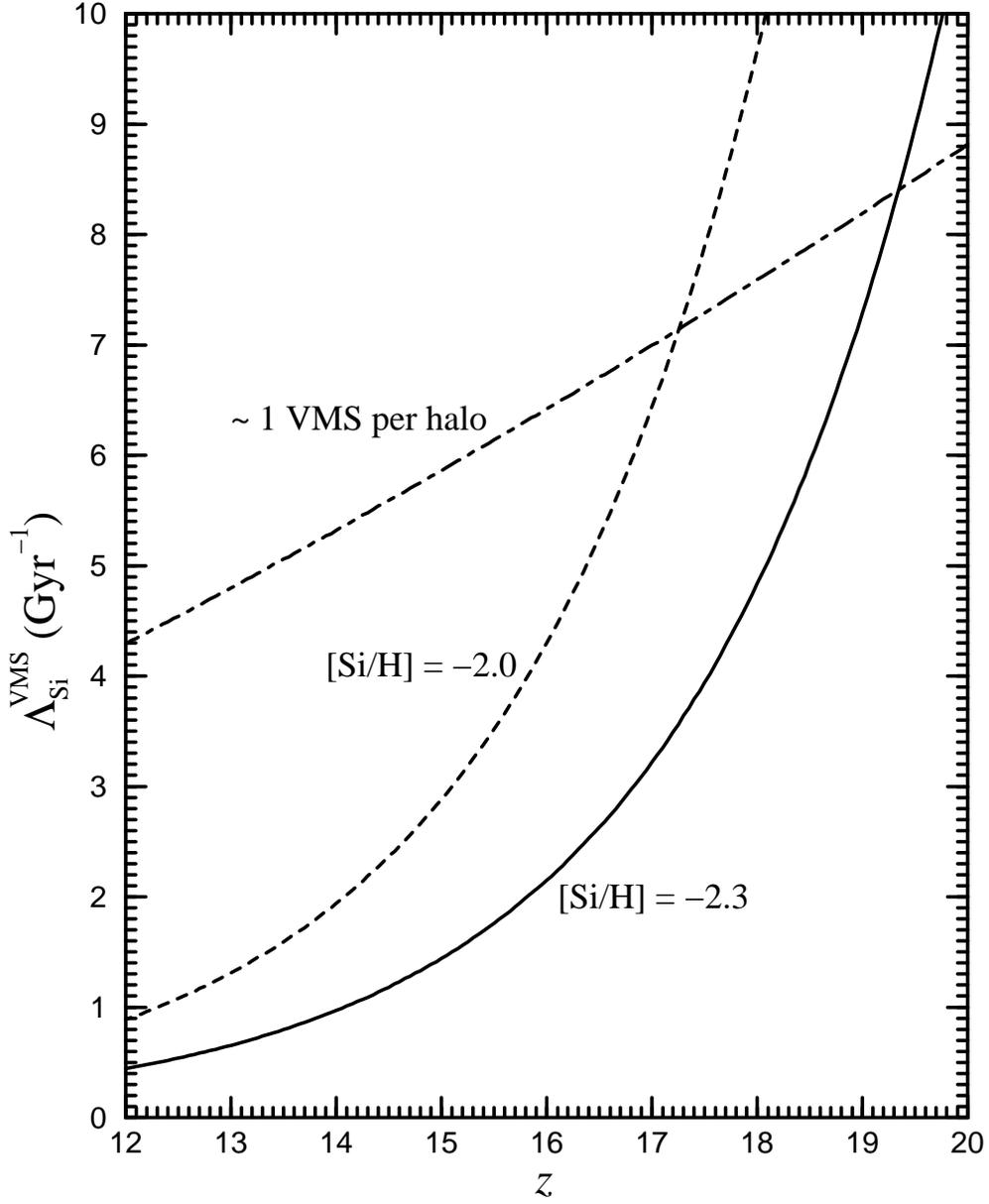}
\caption{Production rates of Si required to provide 50\%
(solid curve) and all (dashed curve) of the IGM inventory of
[Si/H]~$=-2.0$ by a given $z$ as calculated from 
equation~(\protect\ref{zigm}). The dot-dashed curve corresponds to
the condition that every halo of mass 
$M_{n\ \sigma}\sim 10^5\,M_\odot$ would have
$R_{n\ \sigma}t(z)\sim 1$ VMS by a given $z$. 
If VMS formation occurs at $\sim 10$\% of the
time in such halos, the bulk of the Si inventory can only be
obtained at $z\lesssim 15$.}
\end{figure}

\clearpage
\begin{deluxetable}{lccrcccc}
\tablecolumns{8}
\tablewidth{0pc}
\tablecaption{IGM Data, Stellar Models, and Example 
Results\tablenotemark{a}}
\tablehead{
\colhead{}&\colhead{[Si/C]}&\colhead{[C/O]}&\colhead{[O/Fe]}&
\colhead{[Si/H]}&\colhead{[C/H]}&\colhead{[O/H]}&\colhead{[Fe/H]}\\
\colhead{(1)}&\colhead{(2)}&\colhead{(3)}&\colhead{(4)}&\colhead{(5)}&
\colhead{(6)}&\colhead{(7)}&\colhead{(8)}
}
\startdata
IGM&0.74&$-0.50$&\nodata&$-2.0$&$-2.8$&$-2.3$&\nodata\\
VMSs&1.18&$-0.57$&$-0.02$&$-2.3$&$-3.5$&$-2.9$&$-2.9$\\
SNe II&0.42&$-0.42$&0.17&$-2.3$&$-2.7$&$-2.3$&$-2.5$\\
Mixture&0.65&$-0.45$&0.13&$-2.0$&$-2.7$&$-2.2$&$-2.3$\\
\enddata
\tablenotetext{a}{Data on the IGM inventory are for UVB model QG
and are from Schaye et al. 2003,
Aguirre et al. 2004, and Simcoe et al. 2004. Yield ratios for the
VMS model of HW02 and the SN II model of WW95 assuming a Salpeter IMF
(models 2A and 6 in Table 2 of QW05) are given in cols.~(2)--(4). 
The reference solar abundances are the same as adopted in QW05.
A mixture is calculated where VMSs and SNe II 
contribute equally to the Si.
The respective contributions from VMSs and 
SNe II are given in cols.~(5)--(8).}
\end{deluxetable}
\end{document}